\documentclass[prl, twocolumn,superscriptaddress, longbibliography]{revtex4-1}

\usepackage{amssymb}
\usepackage{amsmath}
\usepackage{graphicx}
\usepackage{color}
\usepackage{physics}
\usepackage[colorlinks]{hyperref}

\newcommand{\smx}{\sigma^x}
\newcommand{\smy}{\sigma^y}
\newcommand{\smz}{\sigma^z}

\newcommand{\Cq}{C_{\rm q}}

\newcommand{\Qq}{Q_{\rm q}}

\newcommand{\Lq}{L_{\rm q}}
\newcommand{\Hq}{H_{\rm q}}
\newcommand{\Hr}{H_{\rm r}}
\newcommand{\phiq}{\phi_{\rm q}}

\newcommand{\half}{\frac{1}{2}}
\newcommand{\bphi}{\boldsymbol\phi}

\begin{document}
\title{Optimal gauge for the multimode Rabi model in circuit QED}
\author{Marco Roth}
\affiliation{JARA Institute for Quantum Information (PGI-11), Forschungszentrum J\"ulich, 52428 J\"ulich, Germany}
\affiliation{JARA-Institute for Quantum Information, RWTH Aachen University, 52056 Aachen, Germany}
\author{Fabian Hassler}
\affiliation{JARA-Institute for Quantum Information, RWTH Aachen University, 52056 Aachen, Germany}
\author{David P. DiVincenzo}
\affiliation{JARA Institute for Quantum Information (PGI-11), Forschungszentrum J\"ulich, 52428 J\"ulich, Germany}
\affiliation{JARA-Institute for Quantum Information, RWTH Aachen University, 52056 Aachen, Germany}

\date{April 2019}

\begin{abstract}
In circuit QED, a Rabi model can be derived by truncating the Hilbert space of
the anharmonic qubit coupled to a linear, reactive environment. This truncation
breaks the gauge invariance present in the full Hamiltonian. We analyze the determination of an optimal gauge such that the differences between the
truncated and the full Hamiltonian are minimized. Here, we derive a simple
criterion for the optimal gauge. We find that it is determined by the ratio of
the anharmonicity of the qubit to an averaged environmental frequency. We
demonstrate that the usual choices of flux and charge gauge are not
necessarily the preferred options in the case of multiple resonator modes.
\end{abstract}
\maketitle

Circuit QED \cite{Blais2004, Wallraff2004} is a central subject of quantum
information science that has deepened our understanding of  light-matter
interaction \cite{Schmidt2013, Devoret2013, Wendin2017}. Most implementations
consist of a two-level system (qubit) that is coupled to a linear environment.
The qubit is formed by the two lowest energy levels of an anharmonic
multilevel-system. For the physics of interest only the qubit subspace is
important. The Schrieffer-Wolff (SW) transformation
\cite{Schrieffer1966, Bravyi2011} is the standard method to perturbatively  derive
an effective Hamiltonian description within this subspace.  For most purposes,
it is sufficient to consider the effective Hamiltonian only to first order, yielding the well known quantum Rabi model (QRM). However, since the
Hamiltonian of the non-truncated system is unique only up to a unitary
transformation, the effective description is gauge dependent to every finite
order \cite{Lamb1952,Yang1976}. This gauge ambiguity becomes particularly
important in the (ultra) strong coupling regime. It has been found that the
QRM derived in a gauge where the qubit-resonator coupling is mediated by the
flux variables leads to different predictions than the one where the coupling
is mediated by the charge variables \cite{DeBernardis2018a, DeBernardis2018b,
Stefano2018}.

In this work, we look at the issue from a different perspective. We use the
gauge degree of freedom to find an optimal gauge such that the results of the
effective model are as close as possible to full model. Importantly, we
take account of the need for a multimode description \cite{Nigg2012, Solgun2014, ParraRodriguez2018, Hassler2019} in the quest
for achieving the ultra-strong coupling regime \cite{Gely2017, Bosman2017,
Manucharyan2017, FriskKockum2019}. To increase the flexibility, we not only
consider the extremal cases of purely flux or charge mediated coupling but
perform a general gauge transformation that smoothly interpolates between
the two. A similar transformation has been used in \cite{Stokes2019} to extend
the Jaynes-Cummings model into the ultra-strong coupling regime.

We find that already the second order term of the effective Hamiltonian within the
SW method is a good indicator of the validity of the QRM. Based on this
observation, we derive a simple analytical criterion for the optimal gauge and
benchmark it against numerical simulations of the full problem.
\begin{figure}[t]
	\centering
        \includegraphics[width=\linewidth]{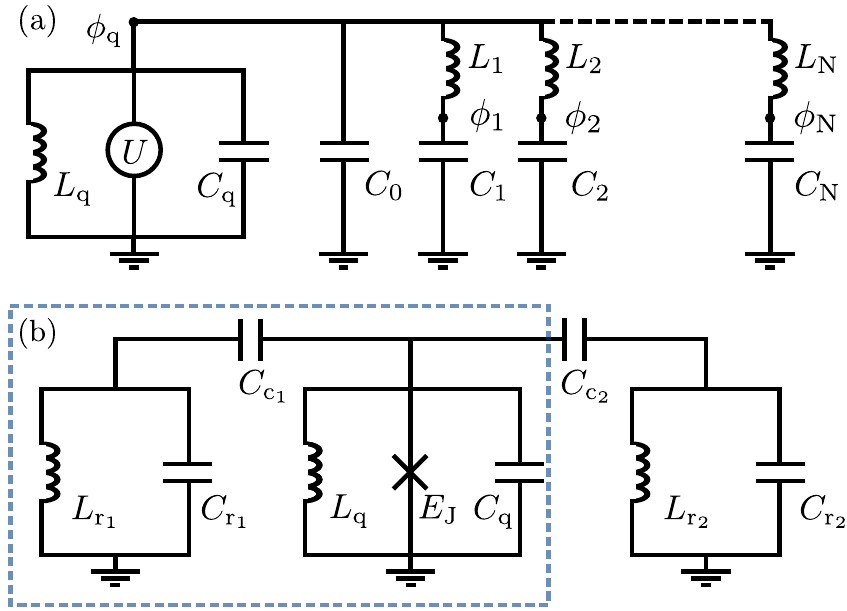}
  \caption{(a) Circuit diagram of a qubit with potential $U(\phiq)$,
  capacitance $\Cq$, and inductance $\Lq$ that is coupled to a general
  reactive environment. In the Forster form, the latter is represented by $N$
  resonators with capacitances $C_k$ and inductances $L_k$ ($k=1,\dots,N$).
  (b) Fluxonium qubit (consisting of a Josephson junction with energy
  $E_\text{J}$ in parallel to the capacitance $\Cq$ and a large inductance
  $\Lq$), capacitively coupled to two resonators  with inductances
  $L_{r_k}$ and capacitances $C_{r_k}$ ($k=1,2$). In the text, the
  example of a single resonator (dashed box) is treated separately.}\label{fig:figure1}
\end{figure}
For a qubit coupled to a single-mode resonator, the flux gauge is always the
best gauge \cite{DeBernardis2018a, DeBernardis2018b,Manucharyan2017}. This
serves as an analogue of the dipole gauge in quantum optics
\cite{CohenTannoudji2007}.  Considering more than one mode drastically changes
this simple picture. The optimal gauge may now deviate from the pure flux gauge
as can be demonstrated with two resonator modes. We show that this
already has implications for weak to moderate coupling.

\textit{General case}.---Consider a qubit consisting of an LC-oscillator in
parallel with a symmetric potential $U(\phi_{\rm q})$ that is coupled to a
linear, reactive environment, cf.\ Fig.~\ref{fig:figure1}(a). We denote the
qubit Hamiltonian $\Hq$ and the resonator Hamiltonian $\Hr$. They are coupled
via the interaction $V$ such that the total Hamiltonian is given by
$H=\Hq+\Hr+V$. Using the unitary freedom of the Hamiltonian formalism, we
introduce a gauge parameter $\eta\in[0,1]$ that linearly interpolates between
a qubit-resonator interaction mediated by the flux variables $\phi_k$ (for
$\eta=0$) and the charge variables $Q_k$ (for $\eta=1$) \cite{Supplement}. We
will refer to these extremal cases as the flux and the charge gauge,
respectively. For a general gauge, the interaction reads
\begin{align}\label{eq:V_eta}
  V(\eta)&=-\sum_{k=1}^N\left[\frac{(1-\eta)\phiq\phi_k}{L_k}+\frac{\eta\,\Qq Q_k}{C_\Sigma}\right]\\
  &\quad\quad\quad+(1-\eta)^2\sum_{k=1}^N\frac{\phi_{\rm
q}^2}{2L_k}+\eta^2\frac{\Bigl(\sum_{k=1}^NQ_k\Bigr)^2}{2C_\Sigma}\,;\nonumber
\end{align}
here, $C_\Sigma=\Cq+C_0$ denotes the total capacitance of the qubit to 
ground. The first term of Eq.~(\ref{eq:V_eta}) is the analogue of the
paramagnetic coupling. The second part is a diamagnetic term that
renormalizes qubit and the resonator frequencies and ensures the gauge
invariance of the full Hamiltonian \cite{Supplement, Malekakhlagh2016}.

For most quantum information applications, we are interested in projecting
$\Hq$ onto a subspace $S=\lbrace \ket{0},\ket{1}\rbrace$ spanned
by the two lowest eigenstates. To obtain an effective Hamiltonian, we apply
the SW method resulting in $H_{\rm eff}=\sum_{j=0}^K H_j$ to $K$-th order.
The first order result $H_0 +H_1$ corresponds to the projection of $H$
onto $S$. It is equivalent to the generalized QRM \cite{Supplement,note_sym}
\begin{align}\label{eq:rabi_model} 
  &H_{\rm QRM}(\eta) = -\frac{\hbar\omega_{10}^{\rm q}}{2}\smz+\sum_{k=1}^N
  \hbar\omega_ka_k^\dag a_k\\
  &\quad\quad+\hbar\sum_{k=1}^N\Bigl[ (1-\eta)g_k^\phi 
    \smx (a_k+a^\dag_k)+\eta
g^Q_k \smy (a_k- a_k^\dag)\Bigr]\,,\nonumber
\end{align}
where $\hbar\omega_{nm}^{\rm q}$ is the energy difference between the $n$-th
and the $m$-th eigenstate of $\Hq$ and $\sigma^j$ ($j=x,y,z$) denote the Pauli
operators. In Eq.~\eqref{eq:rabi_model}, we have rewritten the variables of the $k$-th
resonator mode with frequency $\omega_k$ in terms of bosonic creation
operators $a^\dag_k$ and annihilation operators $a_k$. The coupling between
the qubit and the $k$-th resonator mode is given by
$g_k^\phi=\mel{1}{\phiq}{0}\sqrt{Z_k/2 \hbar L_k^2}$ and
$g_k^Q=\mel{1}{\Qq}{0}/\sqrt{2 \hbar Z_k C_\Sigma^2}$, where $Z_k$ is the
characteristic impedance of the $k$-th mode. In deriving
Eq.~(\ref{eq:rabi_model}), we neglected the diamagnetic shift due to the
second term present in Eq.~(\ref{eq:V_eta}) for simplicity. For weak coupling, the
diamagnetic shift is irrelevant. In general, it can be accounted for
 using symplectic diagonalization \cite{Idel2016, Simon1999}.

Restricting the perturbative series of $H_{\rm eff}$ to any finite order
necessarily results in a gauge dependent model. The source of the gauge
dependence of the QRM is that the coupling between the subspace $S$ and its
orthogonal complement $S^\bot$ is not properly taken into account in the
projection. Increasing the order $K$ weakens the gauge dependence 
\cite{Cederbaum1989, Aharonov1979} at the expense of introducing a dressed
basis that results in a model that strays quite far from the natural
interpretation of the QRM.  In this respect, the lowest order approximation
provided by the QRM is an appealing model as it
yields a low-energy description without rotating the basis. 
In the simple effective model Eq.~(\ref{eq:rabi_model}), choosing a gauge such that the QRM  accurately captures the physics of the full Hamiltonian is crucial. 
We are thus concerned with the task of finding an optimal gauge parameter
$\eta_{*}$ such that the differences between the QRM and the full Hamiltonian
are minimized.

\textit{A criterion for the optimal gauge}.---To address this issue, we note
that the validity of the QRM is directly proportional to the coupling strength
between $S$ and $S^\bot$. The higher order SW terms $H_j$ ($j>1$) can therefore be
used as an estimator for the difference between the full model and its
effective description as a QRM. Based on this observation, we derive an analytic criterion for the optimal gauge.

In particular, we focus on the second order term $H_2$, which will provide the
largest corrections to $H_{\rm QRM}$ for weak coupling. $H_2$ is proportional
to matrix elements $V_{nm}=\mel{n}{V}{m}$ of the interaction, where
$\ket{n}\in S$ and $\ket{m}\in S^\bot$. Motivated by Eq.~(\ref{eq:V_eta}), we
define the paramagnetic flux coupling operator $G^{\phi}_{k}= \phi_{{\rm
q}}\phi_k^{\rm zp}/\hbar L_k$ and the charge coupling operator $G^{Q}_k=
Q_{\rm q}Q_k^{\rm zp}/\hbar C_\Sigma$  \cite{coupling_projection}. Here, we
have approximated the resonator matrix elements by their zero point
fluctuations $\phi_k^{\rm zp}\simeq\sqrt{\hbar Z_k}$ and $Q_k^{\rm
zp}\simeq\sqrt{\hbar/Z_k}$, respectively. In order to estimate the relevance
of the flux versus the charge coupling (for the transition $m\mapsto n$), we
introduce the ratio $f_{nm}=\bigl[\sum_{k}
(G^{\phi}_{k})_{nm}\bigr]/\bigl[\sum_k (G^{Q}_{k})_{nm}\bigr]$. Using the fact
that $(Q_{{\rm q}})_{nm}=i\omega_{nm}^{\rm q}C_\Sigma(\phi_{{\rm q}})_{nm}$,
it can be compactly  rewritten as
\begin{align}
\abs{f_{nm}}
= \frac{\sum_k p_k \omega_k}{\omega^{\rm
q}_{nm}}=\frac{\bar{\omega}}{\omega^{\rm q}_{nm}}\,,\label{eq:f_nm}
\end{align}
where $\bar \omega$ is the average of the resonator frequencies $\omega_k$
with the weights $p_k = Z_{k}^{-1/2} / \bigl(\sum_{l} Z_l^{-1/2}\bigr)$.

The interpretation of Eq.~(\ref{eq:f_nm}) is as follows: if
$\abs{f_{nm}}\ll1$, the coupling between $S$ and $S^\bot$ in the flux gauge is
much smaller than the coupling in the charge gauge. The QRM with
$\eta\approx0$ is therefore a good approximation of the full model, making the
flux gauge the preferred choice. But, if $\abs{f_{nm}}\gg 1$, the
coupling of the qubit subspace to higher levels is small in the charge gauge
which thus is the optimal gauge. In the intermediate regime, where
$\abs{f_{nm}}\simeq 1$, both, flux and charge variables contribute similarly
to the coupling between $S$ and $S^\bot$. Consequently, we expect the optimal
gauge to be neither the pure charge nor the flux gauge but a mixed gauge with
$\eta \neq 0,1$. 

For weak qubit-resonator interactions, the dominant contribution to $H_2$ will
be due to the coupling of the first and second excited level of the qubit. The
character of the coupling of the optimal gauge is therefore mostly determined
by the ratio of the anharmonicity of the qubit to an effective frequency of
the linear environment. We conclude that $f_{21}$ of Eq.~(\ref{eq:f_nm})
provides a simple estimation of the optimal coupling. It requires only
knowledge of the qubit anharmonicity and the frequency and impedances of the
linear environment.  We illustrate these findings with two specific examples
in the following.

\textit{Single resonator.}---First, we consider a qubit coupled to a single
resonator ($N=1$). Note that in this case the average frequency $\bar{\omega}$
in Eq.~(\ref{eq:f_nm}) is equal to $\omega_1$. For the interaction between the
qubit and the resonator mode to be appreciable, we assume that
$\omega_1\simeq\omega^{\rm q}_{\rm 10}$. Consequently, Eq.~\eqref{eq:f_nm}
yields $\abs{f_{21}}\simeq \omega^{\rm q}_{10}/\omega^{\rm q}_{21}$ and the
optimal gauge is solely determined by the properties of the qubit. To
reach strong coupling, the qubit has to be anharmonic with
$\omega_{10}^{\rm q}\ll\omega_{\rm 21}^{\rm q}$ \cite{Manucharyan2017}. This implies 
$\abs{f_{21}}\ll 1$, so we find that the flux gauge is always the optimal
gauge for this case.

To demonstrate this result, we numerically study the fluxonium qubit with
$E_{\rm L} = (\phi_0/2\pi)^2/\Lq \lesssim E_{\rm J}$ and  $U(\phiq) = -E_{\rm
J}\cos[2\pi(\phiq-\phi_{\rm ext})/\phi_0]$ \cite{Manucharyan2009};
\begin{figure}[t]
	\centering
        \includegraphics[width = 8.6cm]{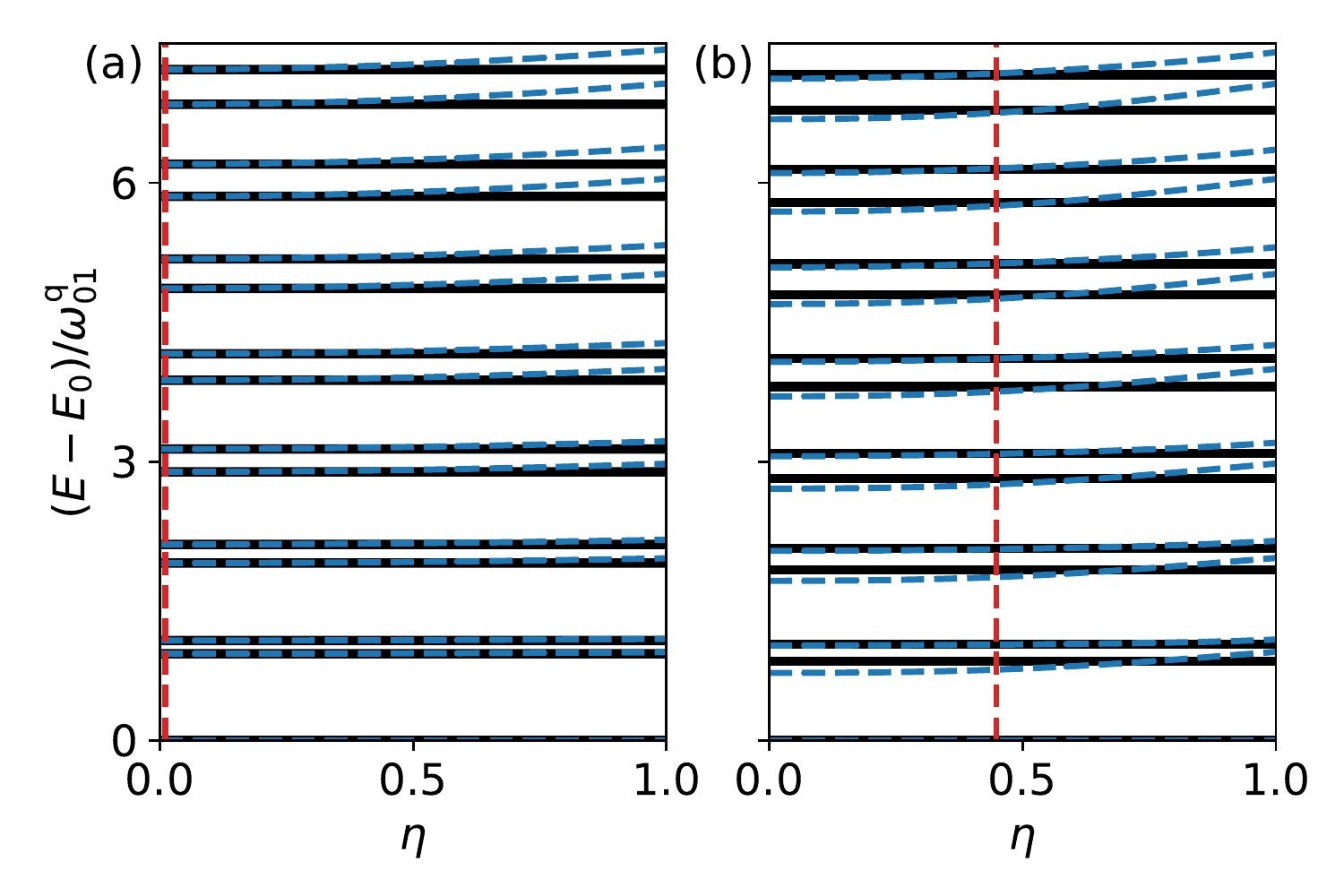}
	\caption{ Spectrum of the full Hamiltonian $H$ (solid lines) and the
	Rabi model Hamiltonian $H_{\rm QRM}$ (dashed lines) for a fluxonium
	qubit coupled to a single resonator (a) and to two resonators (b). The
	qubit parameters are $(E_{\rm J}, E_{C}, E_{L})=
        \hbar(12.5, 3.75,
	0.5)\,\rm GHz$, where $E_C=e^2/2C_\Sigma$ and $E_{\rm L} = (\phi_0/2\pi)^2/\Lq$. The resulting qubit frequency is $\omega_{\rm
	10}^{\rm q}=0.5\,\rm GHz$ and $\omega_{21}^{\rm q}=13\,\rm
	GHz$. Furthermore,  $\omega_1=\omega_{10}^q$ and
	$g^\phi_1/\omega_1=0.07$. In (b) the parameters of the second
	resonator are $C_{\rm r_2}=C_{\rm r_1}$, $C_{\rm c_2}=C_{\rm c_1}$ and
	$\omega_2\approx\omega_{12}^{\rm q}$ such that $\bar{\omega}=10.7\,\rm
	GHz$. The value $\eta_*$ that minimizes $\lVert H_2\rVert_*$ is
	shown as a vertical dashed line.} \label{fig:fluxonium_spectra}
\end{figure}
here, $E_{\rm J}$ is the Josephson energy, $\phi_0=h/2e$ is the
superconducting flux quantum, and $\phi_{\rm ext}$ is an external magnetic
flux threading the superconducting loop. We set the external flux to the
degeneracy point $\phi_{\rm ext}=\tfrac12\phi_0$ which results in a symmetric
potential. The qubit  parameters are chosen such that the qubit is strongly
anharmonic with  $\omega_{21}^{\rm q}/\omega^{\rm q}_{\rm 10}\approx 25$  (see
Fig.~\ref{fig:fluxonium_spectra} for details). The fluxonium qubit is coupled
to a parallel combination of a capacitor $C_{\rm r_1}$ and inductance $L_{\rm
r_1}$ which together form a resonator with a frequency $\omega_1=\omega_{10}^{\rm q}$, cf.\ Fig.~\ref{fig:figure1}(b) (dashed box).
The setup can be mapped to the canonical Foster circuit with $N=1$ shown in
Fig.~\ref{fig:figure1}(a) \cite{Supplement}.

Figure \ref{fig:fluxonium_spectra}(a) shows the spectrum of the full
Hamiltonian (solid) compared to the spectrum of $H_{\rm QRM}$ (dotted) as a
function of $\eta$. The spectra agree well in the flux gauge ($\eta=0$).  For
increasing values of $\eta$, that is for more charge-like gauges, the spectral
agreement between truncated and full model decreases. The disagreement is more
pronounced in levels with higher energy as they are closer to the energy of
the second excited level of the qubit. We observe that $f_{21}$ of
Eq.~(\ref{eq:f_nm}) is suitable for estimating the overall tendency for 
being charge or flux-like. A more quantitative estimate of the optimal coupling $\eta_*$ can be
obtained by calculating the norm of $H_2$. Based on the discussion surrounding
Eq.~(\ref{eq:f_nm}), we expect that $\eta_*$ is approximately the $\eta$ for
which the norm $\lVert H_2\rVert_*$ \cite{Norm} is minimized. For the
parameters in Fig.~\ref{fig:fluxonium_spectra}(a), the minimum of $\lVert
H_2\rVert_*$ is at $\eta=0$, which is shown in red (dotted) and agrees well
with the visual impression conveyed by the spectrum. A quantitative analysis
can be found in Ref.~\cite{Supplement}.

\textit{Two resonators.}--- As a second example, we treat the case where
there are two relevant modes ($N=2$). As before, the first mode is close to
resonance with the qubit frequency.  The second mode with frequency $\omega_2$
can be interpreted as a parasitic mode. Since the average frequency
$\bar{\omega}$ in Eq.~(\ref{eq:f_nm}) is a function of all modes coupled to
the qubit, the optimal gauge is now also dependent on the parasitic mode. This
is true even for strongly off-resonant modes, as the coupling to higher modes in the flux gauge increases $\propto (\omega_2)^2$ at fixed
impedance, see Eq.~\eqref{eq:V_eta}. As a result, for large detuning with
$\omega_2\gg\omega_{21}^{\rm q}$, the charge gauge becomes more favorable. In contrast to the single-mode case, the optimal gauge for two
resonators is not determined by the properties of the qubit alone but depends
on the parameters of the whole circuit.

To show this effect, we perform numerical simulations of the circuit in
Fig.~\ref{fig:figure1}(b). The fluxonium is capacitively coupled to two parallel LC
oscillators via the capacitances $C_{\rm c_1}$ and $C_{\rm c_2}$. This circuit
can be mapped to the canonical Foster circuit depicted in
Fig.~\ref{fig:figure1}(a) \cite{Supplement}.
Figure~\ref{fig:fluxonium_spectra}(b) shows the spectrum of the full
Hamiltonian (black, solid) and the QRM (dashed, blue) as a function of $\eta$.
The parameters of the qubit and the first resonator are the same as in
Fig.~\ref{fig:fluxonium_spectra}(a). The frequency of the second resonator,
however, is significantly larger such that
$\bar{\omega}\approx\omega_{12}^{\rm q}$. In contrast to the single-resonator
case, the spectral lines of $H$ and $H_{\rm QRM}$ do not cross at $\eta
\approx 0$ but rather around $\eta\approx 0.5$, suggesting the optimal gauge does not
coincide with the usual ad-hoc choices of the flux or charge gauge. This is in
agreement with the prediction based on the minimization of $\lVert H_2\rVert_*$ which
yields $\eta_*=0.45$ (shown as a dashed vertical line).

The deviation between $H$ and $H_{\rm QRM}$ is state dependent for finite
qubit anharmonicities, a fact that we have neglected so far. As a result, the
intersection of the spectral lines of $H$ and $H_{\rm QRM}$ in
Fig.~\ref{fig:fluxonium_spectra}(b) is shifted towards smaller values of
$\eta$ for increasing energy of the levels. In the intermediate regime where
$\omega_{\rm q}^{21}\simeq\bar{\omega}$ ($f_{21}\simeq 1$), the optimal gauge
is thus always a compromise which minimizes the differences of $H$ and
$H_{\rm QRM}$ in the relevant spectral range.

To demonstrate the dependence of the optimal gauge on $\bar{\omega}$, we keep
the frequency $\omega_1$ of the first mode in resonance with the qubit while
varying $\bar{\omega}$. To simulate an experimentally feasible scenario, we
choose the inductance $L_{r_2}$ of the second resonator as the parameter that
we vary \cite{Castellanos-Beltran2008}. Decreasing $L_{\rm r_2}$ while keeping
all other parameters constant increases the frequency $\omega_2$ of the
parasitic mode while simultaneously decreasing its impedance. Since
$\bar{\omega}$ decreases with the square root of $Z_2$ but increases linearly
with $\omega_2$, the average frequency $\bar{\omega}$ grows with decreasing
$L_{r_2}$.  To quantify the agreement between the full Hamiltonian and the
QRM, we use the standard deviation $\sigma = \bigl[\sum_{i=0}^M (E_i -
e_i)^2/M\bigr]^{1/2}$ between the energies $E_i$ of the full Hamiltonian and
the energies $e_i$  of the QRM (measured from the respective ground-state
energy). We denote the value of $\eta$ for which $\sigma$ is minimized by
$\eta_{\sigma}$.

\begin{figure}[t]
    \centering
    \includegraphics[width = 8.6cm]{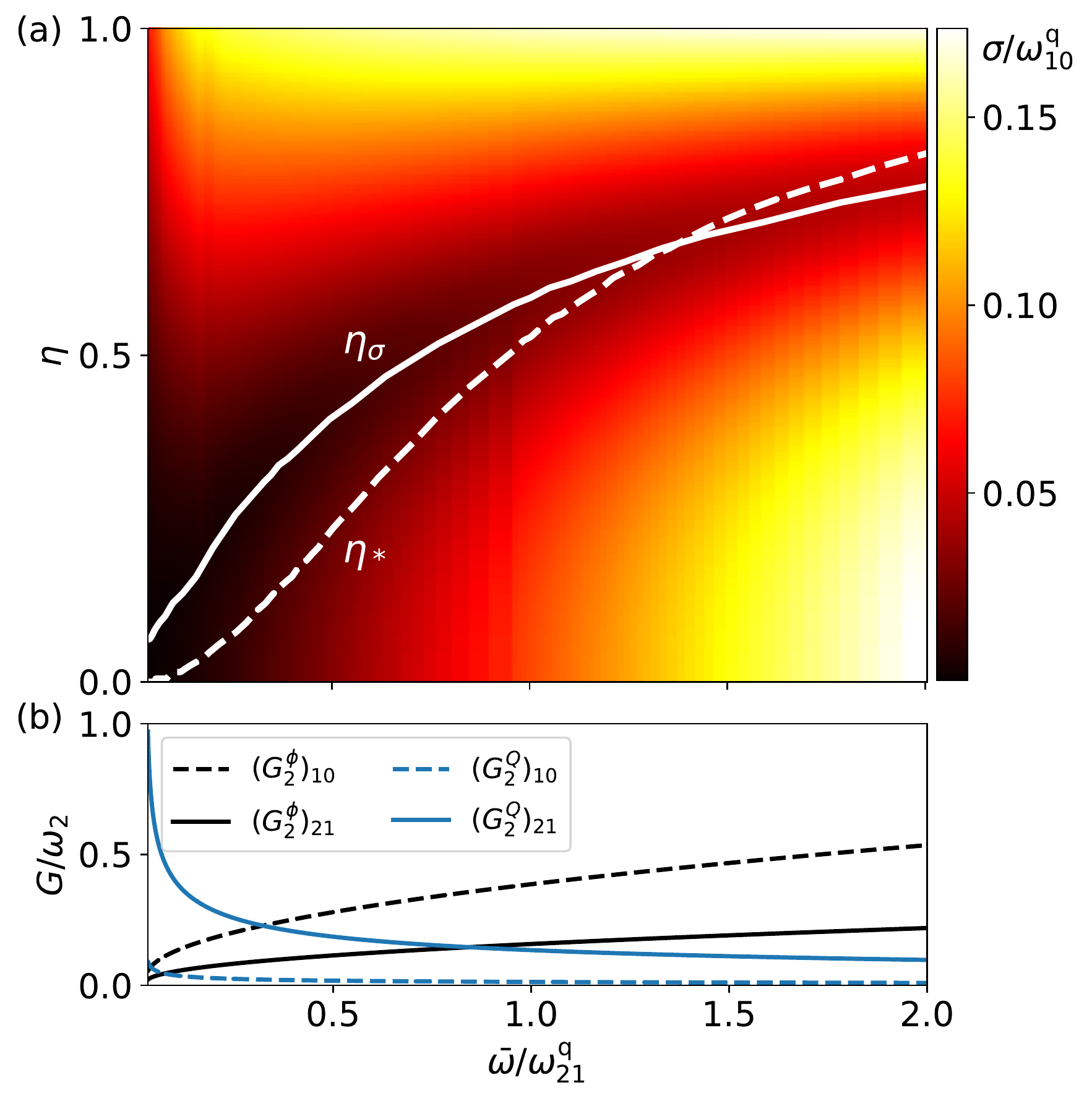}
	\caption{(a) Deviation $\sigma$ of the energy eigenvalues of the
	full model from the energy eigenvalues of the QRM as function of $\eta$ and
	$\bar{\omega}$. The qubit parameters and the parameters of the first
	resonator are the same as in Fig.~\ref{fig:fluxonium_spectra}. The coupling
	capacitances to both resonators are equal $C_{c_1}=C_{c_2}=C_c$. The
	average frequency $\bar{\omega}$ is varied by changing the
	inductance of the second resonator $L_{\rm r_2}$. The value
	$\eta_\sigma$ for which $\sigma$ is minimized is shown in as a solid line. The value of $\eta_*$ for which $\lVert H_2\rVert_*$ is
	minimized is shown as a dashed line. (b) Coupling strength between
	qubit levels $n$ and $m$ (see legend) as a function of $\bar{\omega}$.}
\label{fig:eta_vs_omega}
\end{figure}
Figure~\ref{fig:eta_vs_omega}(a) shows $\sigma$ as a function of
$\bar{\omega}$ and $\eta$ for the first 15 energy levels. We see that $\eta_\sigma \approx 0$
(flux gauge) for  $\bar{\omega}\ll\omega_{10}^{\rm q}$. Increasing the average
frequency $\bar{\omega}$, the optimal gauge moves towards the charge gauge.
Furthermore, we note that although the minimal value of $\sigma$ increases
with increasing $\bar{\omega}$, the overall deviation between the full model and the QRM at $\eta_{\sigma}$ is only a few percent. The value $\eta_*$ which minimizes $\lVert H_2\rVert_*$ is shown as a dashed line. It can be observed that $\eta_*$ behaves similarly to $\eta_\sigma$.

To support our discussion surrounding Eq.~(\ref{eq:f_nm}), we analyze the
coupling between $S$ and $S^\bot$. Figure~\ref{fig:eta_vs_omega}(b) shows
$(G^{\phi}_{2})_{nm}$ (black) and $(G^{Q}_{2})_{nm}$ (blue) for the parameters
of Fig.~\ref{fig:eta_vs_omega}(a). In general, the charge coupling $G^Q$
decreases while the  flux coupling $G^\phi$ increases with increasing
$\bar{\omega}$. For small values of $\bar{\omega}$, the dominant quantity
is $(G^{Q}_{2})_{21}$. This results in a large coupling between $S$ and
$S^\bot$ in the charge gauge, making the flux gauge the preferred choice. As
$\bar{\omega}$ increases, the coupling to the higher qubit levels in the
charge variables decreases and eventually becomes comparable to the coupling
in the flux variables, making the choice of the optimal gauge less trivial.

\textit{Conclusion}.--- We have analyzed the gauge dependence of the effective
description of an anharmonic system coupled to a general linear environment.
Using a SW transformation, we have derived a simple, analytic criterion that predicts the
optimal gauge where the physics of the non-truncated Hamiltonian is accurately
captured by the QRM. We have demonstrated that the optimal gauge for a qubit resonantly coupled to a single resonator is completely determined by the qubit parameters and is in the flux-like regime for strongly anharmonic qubits. We have seen that coupling a qubit to more than one mode can result in an optimal gauge that is neither the charge nor the flux gauge but a non-trivial combination of the two. This is especially relevant with the increasing interest in the ultra-strong coupling regime which raises the need for multimode descriptions.

\bibliography{literature}

\begin{thebibliography}{35}
\bibitem [{not()}]{note_sym}%
  \BibitemOpen
  \href@noop {} {}\bibinfo {note} {Here, the parity symmetry $U(-\phiq) =
  U(\phiq)$ is important for giving the selection rules $\langle j |\Qq |j
  \rangle = \langle j |\phiq |j \rangle = 0$.}\BibitemShut {Stop}%
\bibitem [{cou()}]{coupling_projection}%
  \BibitemOpen
  \href@noop {} {}\bibinfo {note} {Note that the couplings $G^\phi$ and $G^Q$
  are generalizations of the couplings introduced in Eq.~(\ref{eq:rabi_model})
  such that $\mel{0}{G^\phi}{1}=-g^\phi$ and
  $\mel{0}{G^Q}{1}=-ig^Q$.}\BibitemShut {Stop}%
\bibitem [{Nor()}]{Norm}%
  \BibitemOpen
  \href@noop {} {}\bibinfo {note} {In Figs.~\ref{fig:fluxonium_spectra} and
  \ref{fig:eta_vs_omega} we have used the trace norm
  $\lVert\mathcal{H}\rVert_*=\sum_k\abs{\lambda_k}$ for an Hermitian operator
  $\mathcal{H}$ with eigenvalues $\lambda_k$. Since we are only interested in
  the value of $\eta$ that minimizes the norm of $H_2$, other norms can be used
  as well.}\BibitemShut {Stop}%
\end{thebibliography}
\onecolumngrid
\clearpage
\setcounter{equation}{0}
\renewcommand\theequation{S\arabic{equation}}
\setcounter{figure}{0}
\renewcommand\thefigure{S\arabic{figure}}
\section*{Supplement}\setcounter{page}{1}
\section{Gauge transformation}
\label{sec:gauge_transformation}
In this chapter, we introduce the gauge transformation discussed in the main text on a Lagrangian level. The Lagrangian $\mathcal{L}(\bphi,\dot{\bphi})$ of a qubit in potential $U$ coupled to a general linear environment is a function of the fluxes $\bphi=\left(\phiq,\phi_1,\dots,\phi_{\rm N}\right)^T$ and the voltages proportional to $\dot{\bphi}$. Here, the dot denotes the time derivative. Using the capacitance matrix $C$ and the inverse of the inductance matrix $M=L^{-1}$, it can be written as
\begin{equation}
\mathcal{L}(\bphi,\dot{\bphi})=\half\dot{\bphi}^TC\dot{\bphi}-\half\bphi^TM\bphi-U(\phiq)\,.\label{eq:lagrangian}
\end{equation}
The Euler-Lagrange equations are invariant under coordinate transformations. The choice of coordinates corresponds to choosing a specific gauge in electromagnetic field theory. In Eq.~(\ref{eq:lagrangian}), the flux variable $\phiq$ is distinguished from the rest by the presence of the potential $U(\phiq)$. We thus consider coordinate transformations $\phi=T\phi'$ that preserve this structure and leave the variable $\phiq$ invariant. In its most general form, such a transformation is given by
\begin{equation}
T=
\begin{pmatrix} 
1 & {\bf 0} \\
\bf{t} & R
\end{pmatrix}\,,\label{eq:gauge_transformation}
\end{equation}
where $R$ is an invertible matrix and $\bf{t}$ is an $N$-dimensional vector.

 In general, both $C$ and $M$ provide a qubit-resonator coupling. In the following, we show that it is not possible to decouple the qubit from the resonator with a transformation of the form of Eq.~(\ref{eq:gauge_transformation}). To demonstrate this, we write $C$ and $M$ in the same block structure as Eq.~(\ref{eq:gauge_transformation})
\begin{equation}
C=
\begin{pmatrix} 
\kappa & {\bf c}^T \\
\bf{c} & C_{\rm r} 
\end{pmatrix}\,,\quad
M=
\begin{pmatrix} 
\mu & {\bf m}^T \\
\bf{m} & M_{\rm r} 
\end{pmatrix}\,.
\end{equation}
Here, $\kappa$ is the qubit capacitance and $\mu$ is the qubit inductance. Moreover, $C_{\rm r}$ and $M_{\rm r}$ are the capacitance and inverse inductance matrices of the resonators. The vectors $\bf{c}$ and $\bf{m}$ couple the flux and voltage variables of the qubit and the resonators. Under the transformation Eq.~(\ref{eq:gauge_transformation}), $C$ and $M$ transform as $C'=T^TCT$ and $M'=T^TMT$ which yields the transformed off-diagonal blocks ${\bf c'}=R{\bf c}+RC_{\rm r}{\bf t}$ and ${\bf m'} = R{\bf m}+RM_{\rm r}{\bf t}$. Therefore, in order for ${\bf c'}$ and ${\bf m'}$ to vanish at the same time, the following equations have to be satisfied
\begin{align}
{\bf m} - M_{\rm r}C_{\rm r}^{-1} {\bf c} &=0\,,\\
{\bf c}-C_{\rm r} M^{-1}_{\rm r} {\bf m} & =0\,.
\end{align}
These equations can only be satisfied simultaneously if the qubit and the resonators are uncoupled. Nevertheless, one can choose coordinates such that the qubit is coupled to the resonator only through the capacitance or the inverse inductance matrix, respectively. In the following, we fix ${\bf t}$ and introduce a gauge parameter $\eta$ that linearly interpolates between these two extreme cases
\begin{equation}
{\bf t}=-\left[(1-\eta)C_{\rm r}^{-1}{\bf c}+\eta M_{\rm r}^{-1}{\bf m}\right]\,.\label{eq:t_gauge}
\end{equation}
One can easily verify that $\eta=0$ results in a block-diagonal capacitance matrix $C'$. The coupling is then completely inductive and we call the corresponding gauge \emph{flux gauge}. On the other hand, $\eta=1$ block-diagonalizes $M'$ which results in a purely capacitive coupling. We call the corresponding gauge \emph{charge gauge}.
\section{Full Hamiltonian}
\label{sec:full_hamiltonian}
Figure~\ref{fig:figure1}(a) in the main text shows a qubit in a potential $U$ coupled to a general admittance modelled by a series of LC oscillators. Choosing the flux gauge to represent the circuit (determined by the choice of ground node \cite{Devoret1995}), the Lagrangian is given by
\begin{equation}
\mathcal{L}(\bphi,\dot{\bphi}) = \frac{C_\Sigma\dot{\phi}_{\rm q}^2}{2}-\frac{\phi_{\rm q}^2}{2\Lq}-U(\phi_{\rm q})+\sum_{k=1}^N\left[\frac{C_k\dot{\phi}_k^2}{2}-\frac{\left(\phi_k-\phi_{\rm q}\right)^2}{2L_k}\right]\,.\label{eq:flux_gauge_lagrangian}
\end{equation}
Here, $C_\Sigma=\Cq + C_0$ is the total capacitance of the qubit to ground. We introduce a gauge parameter $\eta$ by performing the variable transformation Eq.~(\ref{eq:gauge_transformation}) discussed in the previous  section. We use the specific ${\bf t}$ from Eq.~(\ref{eq:t_gauge}). For the Lagrangian in Eq.~\eqref{eq:flux_gauge_lagrangian} the coupling vectors read ${\bf c}={\bf 0}$ and ${\bf m}=(-L^{-1}_1,-L^{-1}_2,\dots,-L^{-1}_{\rm N})^T$. The capacitance and inductance matrices of the resonators are diagonal. They are given by $C_{\rm r}=\text{diag}(C_1,C_2,\dots,C_{\rm N})$ and $M_{\rm r}=\text{diag}(L^{-1}_1,L^{-1}_2,\dots,L^{-1}_{\rm N})$. Performing the transformation yields
\begin{equation}
\mathcal{L}'(\bphi',\dot{\bphi}') = \frac{C_\Sigma\dot{\phi}_{\rm q}'^2}{2}-\frac{\phi_{\rm q}'^2}{2\Lq}-U(\phi_{\rm q}')+\sum_{k=1}^N\left[\frac{C_k\left(\dot{\phi}_k'+\eta\dot{\phi}_{\rm q}'\right)^2}{2}-\frac{\left(\phi_k'-\left( 1-\eta\right)\phi_{\rm q}'\right)^2}{2L_k}\right]\,,
\end{equation}
where, $\bphi=T\bphi'$. We define the conjugate momenta $Q_i'=\pdv{\mathcal{L'}}{\phi_i'}$ of the flux variables $\bphi'$, and perform a Legendre transformation which yields the Hamiltonian $H=\sum_iQ'_i\dot{\phi}'_i-\mathcal{L'}$. To obtain  a quantum mechanical description, we promote the canonical variables to operators $\phi'_i\rightarrow\hat{\phi}_i$ and $Q_i'\rightarrow\hat{Q}_i$, and impose the canonical commutation relation $\comm{\hat{\phi}_i}{\hat{Q}_j}=i\hbar\delta_{ij}$, where $\delta_{ij}$ is the Kronecker delta. The total Hamiltonian $H(\eta)=\Hq+\Hr+V(\eta)$ can then be split into a qubit Hamiltonian $\Hq$, a resonator Hamiltonian $\Hr$ and the interaction $V$ with
\begin{subequations}
\begin{align}
H_{\rm q} &= \frac{\hat{Q}_{\rm q}^2}{2C_\Sigma}+\frac{\hat{\phi}_{\rm q}^2}{2\Lq}+U(\hat{\phi}_{\rm q})\,,\qquad H_{\rm r} = \sum_{k=1}^N\frac{\hat{Q}_k^2}{2C_k}+\frac{\hat{\phi}_k^2}{2L_k}\,,\\
V(\eta)&=-\sum_{k=1}^N\left[\frac{(1-\eta)\hat{\phi}_{\rm q}\hat{\phi}_k}{L_k}+\frac{\eta\hat{Q}_{\rm q} \hat{Q}_k}{C_\Sigma}\right]+(1-\eta)^2\sum_{k=1}^N\frac{\hat{\phi}_{\rm q}^2}{2L_k}+\eta^2\frac{\left(\sum_{k=1}^N\hat{Q}_k\right)^2}{2C_\Sigma}\,.\label{eq:V_eta_supp}
\end{align}
\label{eq:H_eta}
\end{subequations}
The interaction $V$ in Eq.~(\ref{eq:V_eta_supp}) is given in
Eq.~(\ref{eq:V_eta}) of the main text where the hats over the operators have been
omitted. Note that the Hamiltonian $H(\eta)$ is related to the Hamiltonian $H(\eta')$ through the unitary
transformation $R=\exp[-i(\eta'-\eta)\hat{\phi}_{\rm q}\sum_k\hat{Q}_k/\hbar]$
such that $R^\dag H(\eta)R=H(\eta')$. The difference of $H(\eta)-H(0)$
corresponds to a pseudopertubation of Ref.~\cite{Aharonov1979}.
\section{Schrieffer-Wolff transformation}
\label{sec:SWT}
In this section, we perform a SW transformation to derive the QRM Eq.~(\ref{eq:rabi_model}).
Similar to the main text, we define the low energy subspace of the qubit
$S=\left\lbrace\ket{0},\ket{1}\right\rbrace$ and its orthogonal complement $S^\bot$.
Furthermore, we define the projector $P=\dyad{0}{0}+\dyad{1}{1}$ onto $S$. The
projector onto $S^\bot$ is then given by $Q=1-P$. The
coupling between the subspaces $S$ and $S^\bot$ is provided by $PVQ$. Performing a SW
transformation to block-diagonalize $H$ with respect to $S$ and $S^\bot$ results in an effective Hamiltonian $H_{\rm eff}=\sum_{j=0}^K H_j$ \cite{Winkler2003}. The zeroth order is given by the projection of the uncoupled Hamiltonian onto $S$
\begin{equation}
H_0=P\Hq P + \Hr=-\hbar\frac{\omega_{10}^{\rm q}}{2}\smz+\hbar\sum_{k=1}^N\omega_ka^\dag_ka_k\,.
\end{equation}
Here, $\hbar\omega_{10}^{\rm q}$ is the energy difference between the ground state and the first excited state of the qubit. Furthermore, we have defined the frequencies $\omega_k=1/\sqrt{L_kC_k}$ and the bosonic raising and lowering operators of the $k$-th mode
\begin{align}
\hat{\phi}_n&=\sqrt{\frac{\hbar Z_k}{2}}\left(a^\dag_k+a_k\right)\,,\\
\hat{Q}_k&=i\sqrt{\frac{\hbar}{2Z_k}}\left(a^\dag_k-a_k\right)\,,
\end{align}
where $Z_k=\sqrt{L_k/C_k}$ is the characteristic impedance of the $k$-th mode. The next order is given by the projection of the interaction $V$ onto $S$
\begin{equation}
H_1 = PV(\eta)P = \hbar\sum_{k=1}^N\Bigl[ (1-\eta)g_k^\phi 
\smx (a_k+a^\dag_k)+\eta
g^Q_k \smy (a_k- a_k^\dag)\Bigr] - \frac{\left(1-\eta\right)\alpha}{2}\smz-\frac{\eta^2\hbar}{2C_\Sigma}\left(\sum_{k=1}^N\frac{ a^\dag_k-a_k}{\sqrt{Z_k}}\right)^2\,,\label{eq:H1}
\end{equation}
where $g_k^\phi=\mel{1}{\phiq}{0}\sqrt{Z_k/2 \hbar L_k^2}$ and
$g_k^Q=\mel{1}{\Qq}{0}/\sqrt{2 \hbar Z_k C_\Sigma^2}$. Furthermore,
$\alpha=(\mel{1}{\phiq^2}{1}-\mel{0}{\phiq^2}{0})\sum_n1/L_n$. The last two
terms in Eq.~(\ref{eq:H1}) are diamagnetic renormalizations of the qubit and
resonator frequencies. If these terms are omitted, the first order effective
Hamiltonian $H_0+H_1$ is equal to the QRM Eq.~(\ref{eq:rabi_model}) in the main text. For weak qubit-resonator coupling this is a reasonable assumption.
\section{Foster Representation}
\label{sec:foster_mapping}
The circuit shown in Fig.~\ref{fig:figure1}(b) can be mapped onto the general Foster form of Fig.~\ref{fig:figure1}(a). Assuming a symmetric coupling $C_{c_1}=C_{c_2}=C_c$, the capacitances and inductances are given by the following substitutions
\begin{align}
C_k = \frac{C_{c}^2}{C_{c}+C_{r_k}}\,\qquad L_k = \frac{L_{r_k} (C_{c}+C_{r_i})^2}{C_{c}^2}\,,
\end{align}
with 
\begin{equation}
C_0= \frac{C_cC_{r_1}}{C_c+C_1}\,,
\end{equation}
for the one-mode setup (dashed box) and
\begin{equation}
C_0=C_c \left(\frac{C_{r_1}}{C_c+C_{r_1}}+\frac{C_{r_2}}{C_c+C_{r_2}}\right)\,,
\end{equation}
for the two-mode setup.

\section{Single Resonator results, detailed}
\label{sec:one_resonator}
\begin{figure}[h]
	\centering
		\includegraphics[width=6 cm]{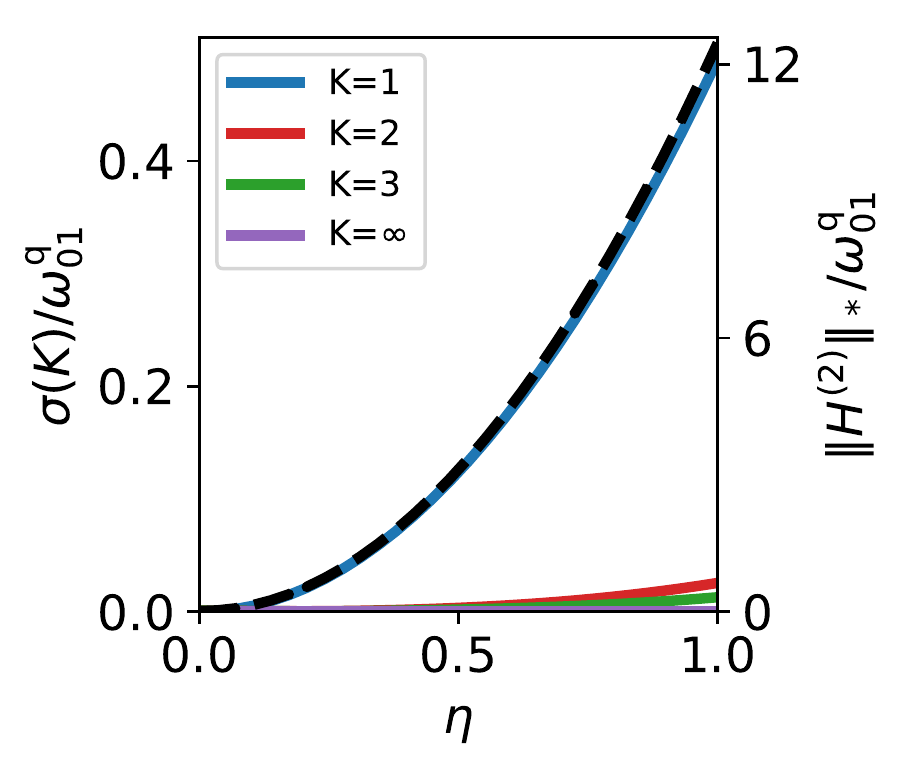}
	\caption{Standard deviation $\sigma$ of the full spectrum and the
        effective Hamiltonian $H_{\rm eff}$ to $K$-th order. The QRM
      corresponds to $K=1$ (blue). Moreover $K=2$ (red), $K=3$ (green) and the
    exact SW transformed Hamiltonian $K=\infty$ (purple) are shown.
  Additionally, the norm $\lVert H_2\rVert_*$ of the first perturbative correction to $H_{\rm QRM}$ is shown (black, dashed). The parameters are the same as in Fig.~\ref{fig:fluxonium_spectra} in the main text.}
	\label{fig:metric_1r}
\end{figure}
In this section, we show supporting data for one qubit coupled to one
resonator. Figure~\ref{fig:metric_1r} shows the standard deviation $\sigma(K) =
\sqrt{(1/M)\sum_{i=0}^M (E_i - e_i(K))^2}$ for the first $15$ states. Note that here
$e_i(K)$ denote the eigenvalues of the SW transformed Hamiltonian $H_{\rm eff}$
to $K$-th order. In Fig.~\ref{fig:metric_1r}, the parameters are the same as in
Fig.~\ref{fig:fluxonium_spectra} in the main text. For all finite values of $K$, the minimum of $\sigma$ is at $\eta\approx 0$ demonstrating that the flux gauge is optimal in this case.

 Furthermore, we see that adding higher order terms to the Rabi
Hamiltonian mitigates the effect of the broken gauge invariance. The deviation between full and effective model becomes less sensitive to variations
in $\eta$ with increasing order in the SW method. The exact SW transformation \cite{Cederbaum1989} (purple) results in a
gauge invariant two-level description. Additionally, 
 the norm of $H_2$ is shown(black, dashed). We observe a non-linear increase towards
charge-like gauges as expected from the previous discussion. For $K=1$ (blue,
solid), we see a strikingly similar functional dependence on $\eta$ as in
$\lVert{H_2}\rVert_*$, indicating that a large part of the corrections to $H_{\rm QRM}$ are already captured by $H_2$.
\end{document}